\documentclass[aip,pop,reprint,showpacs,superscriptaddress]{revtex4-1}
\usepackage{graphicx}
\usepackage{dcolumn}
\usepackage{bm}
\begin{document}


\title{The breather like penetration of the ultra-short linearly polarized laser into over-dense plasmas}
\author{Dong Wu}
\email{wudongphysics@gmail.com}
\affiliation{Center for Applied Physics and Technology, \\Peking University, Beijing, 100871, China.}
\affiliation{Key Laboratory of High Energy Density Physics Simulation, \\Ministry of Education,
Peking University, Beijing, 100871, China. }
\author{C. Y. Zheng}
\email{zheng\_chunyang@iapcm.ac.cn}
\affiliation{Center for Applied Physics and Technology, \\Peking University, Beijing, 100871, China.}
\affiliation{Key Laboratory of High Energy Density Physics Simulation, \\Ministry of Education,
Peking University, Beijing, 100871, China. }
\affiliation{Institute of Applied Physics and Computational Mathematics, \\Beijing, 100088, China.}
\author{X. T. He}
\email{xthe@iapcm.ac.cn}
\affiliation{Center for Applied Physics and Technology, \\Peking University, Beijing, 100871, China.}
\affiliation{Key Laboratory of High Energy Density Physics Simulation, \\Ministry of Education,
Peking University, Beijing, 100871, China. }
\affiliation{Institute of Applied Physics and Computational Mathematics, \\Beijing, 100088, China.}
\date{\today}
\begin{abstract}
Relativistic electromagnetic penetration behavior is reexamined in the relativistic transparency region. 
The interaction is modeled by the relativistic hydrodynamic equations coupled with the full system of Maxwell equations, 
which are solved by a fully implicit energy-conserving numerical scheme. 
For the first time, 
we have studied the penetration behavior through ultra-short circularly polarized and linearly polarized laser pulses interaction with over-dense plasmas. 
It is shown that for the ultra-short circularly polarized laser penetration occurs through a soliton like behavior, which is quite consistent with the existing studies. However, we have found that the ultra-short linearly polarized laser penetrates through a breather like behavior, and there is an energy transition mechanism with a frequency $2\omega_0$ between the penetrated breather like structure and the background plasmas. 
A qualitative interpretation has been given to describe this mechanism of energy exchange. 
\end{abstract}
\maketitle

The dynamics of the penetration of ultra-intense laser irradiation into classically-forbidden plasmas has attracted wide attention for its distinct prospective applications. The penetration of intense laser radiation into over-dense plasma core would be crucial in the concept of the fast ignition fusion\cite{PhysRevLett.86.436}, plasma lens for ultra-intense laser focusing \cite{PhysRevLett.107.265002}and betatron oscillations for high intense electron acceleration\cite{PhyPla.6.2847,lasptbeams.26.51}. Kaw et al. are the first to predict that relativistically strong laser can reduce the effective plasma frequency below the frequency of the laser radiation, allowing the laser to penetrate through the over-dense plasmas\cite{PhyFlu.13.472}. This problem has been greatly enriched by a series of theoretical and particle-in-cell simulation studies\cite{PhysRevLett.69.1383,PhyPla.1.745,PhysRevLett.74.2002,Phys.Rev.E.55.1011,
PhysRevLett.87.185004,PhysRevLett.76.3975,PhyPla.3.2693,Phys.Rev.E.54.1870,Phys.Rev.E.58.4937,
Phys.Rev.E.65.016405,Phys.Rev.E.70.036403,Phys.Rev.A.46.6608,Phys.Rev.E.85.026405,PhysRevLett.103.215005,
Phys.Rev.E.86.056404,PhyPla.17.043102}. 
Tushentsov et al. are the first to study the irradiation penetration of the circularly polarized (CP) laser pulse through fluid-Maxwell simulations\cite{PhysRevLett.87.275002}, but there is a serious shortcoming: the Maxwell equations are solved within the framework of parabolic approximation, which will certainly break down for ultra-short laser pulse.
Afterward, Berezhiani et al. abandon the parabolic approximation and solve the full system of relativistic hydrodynamic and Maxwell equations\cite{PhyPla.12.062308}. However in their results, we have not found the intrinsic application to the study of ultra-short laser penetration into over-dense plasmas. In their fluid-Maxwell simulations, the laser duration and solitons formation time are over $100T_0$. Their simulation results just reconfirm that of Tushentsov et al. with parabolic approximation. It should be emphasized that their is no theoretical analysis on the penetration into over-dense plasmas with linearly polarized (LP) laser, because the second-harmonic oscillating component of the ponderomotive force could disturb the background plasma density and make the problem even more complicated. The difficulty of the fluid-Maxwell simulations in studying LP solitons is that the explicit numerical scheme will break down in dealing with density ripples caused by second-harmonic oscillating ponderomotive force \cite{PhyPla.13.092302}. However, the crucial issue, which is of fundamental influence to the fluid-Maxwell simulations, is the vacuum heating and $\bm{J}\times\bm{B}$ heating mechanisms on the vacuum-plasma interface. These heatings are totally kinetic mechanisms which can not be described by a fluid treatment. Whereas in our work, these strait problems are solved in two ways. A fully implicit energy-conserving numerical scheme is applied to well deal with the density ripples caused by the oscillating term. Abandoning the parabolic approximation \cite{PhyPla.12.062308,PhyPla.13.092302}, we solve the full system of relativistic hydrodynamic and Maxwell equations, and focus on ultra-short laser irradiation with full-width-half-maximum $5T_0$. In such cases the vacuum heating and $\bm{J}\times\bm{B}$ heating can be neglected considering the ultra-short interaction time between the laser and plasmas on vacuum-plasma interface.

In this paper, the interaction between the ultra-short laser irradiation and the plasmas
is modeled by the relativistic hydrodynamic equations coupled with the full system of Maxwell equations, 
which are solved by a fully implicit energy-conserving numerical scheme. 
The penetration of the ultra-short CP irradiation occurs through a soliton like behavior which is consistent with previous theoretical and fluid-Maxwell simulation studies\cite{Phys.Rev.E.62.1234,PhysRevLett.87.275002,PhyPla.12.062308}, and its propagation in the homogeneous background plasmas is quasi-static.  
However, it is shown that, for the ultra-short CP laser, the penetration occurs through a breaker like behavior, and there is an energy transition mechanism with a frequency $2\omega_0$ between the penetrated breather like structure and the background plasmas. 
We have proposed a qualitative interpretation to describe this mechanism of energy exchange. 

The governing equations are the relativistic hydrodynamic fluid equations coupled with the full system of Maxwell equations. 
As we are considering the relativistic intense laser irradiation and the electron quiver velocity under the laser electric field is much greater than the thermal velocity, we treat the plasmas as cold liquid. Below are the group of normalized full system of equations \cite{PhyPla.12.062308,Phys.Rev.E.62.1234}:
\begin{eqnarray}\label{1,2,3,4,5,6,7,8,9} 
&& \frac{\partial^2 a_y}{\partial t^2}-\frac{\partial^2 a_y}{\partial x^2}+(\frac{n_e}{\gamma_e}+Z^2\frac{n_i}{m\gamma_i})a_y=0, \\ 
&& \frac{\partial^2 a_z}{\partial t^2}-\frac{\partial^2 a_z}{\partial x^2}+(\frac{n_e}{\gamma_e}+Z^2\frac{n_i}{m\gamma_i})a_z=0, \\
&& \frac{\partial n_e}{\partial t}+\frac{\partial}{\partial x}(\frac{n_e p_e}{\gamma_e})=0, \\
&& \frac{\partial n_i}{\partial t}+\frac{\partial}{\partial x}(\frac{n_i p_i}{\gamma_i})=0, \\
&& \frac{\partial p_e}{\partial t}+\frac{\partial \gamma_e}{\partial x}-\frac{\partial \phi}{\partial x}=0, \\
&& \frac{\partial p_i}{\partial t}+\frac{\partial \gamma_i}{\partial x}+\frac{Z}{m}\frac{\partial \phi}{\partial x}=0, \\
&& \frac{\partial^2 \phi}{\partial x^2}=(n_e-n_i), \\
&& \gamma_e=[1+a_y^2+a_z^2+p_e^2]^{1/2}, \\
&& \gamma_i=[1+(a_y/m)^2+(a_z/m)^2+p_i^2]^{1/2}.
\end{eqnarray}
The group of equations are normalized in terms of dimensionless quantities $t=\omega_0t$, $x=\omega_0x/c$, 
$a_y=eA_y/m_ec^2$, $a_z=eA_y/m_ec^2$, $n_e=n_e/n_c$, $n_i=n_i/n_c$, $p_e=p_e/m_ec$, $p_i=p_i/m_ic$, $m=m_i/m_e$ and $\phi=e\phi/m_ec^2$, where $\omega_0$ is the laser frequency, $a_y$ and $a_z$ are the normalized amplitude of the laser irradiation, $n_e$ and $n_i$ are the normalized plasma density for electrons and ions, $p_e$ and $p_i$ are the normalized momentum along the laser propagation direction for electrons and ions, $m$ is the normalized ion mass, $\phi$ is the normalized electrostatic potential and $n_c$ is the critical density.

To solve the group of equations, a fully implicit energy-conversing numerical scheme has been applied \cite{numerical}, which can well deal with the density ripples and restrain overshooting of the numerical scheme. Here we have assumed that the boundary conditions [$a_y=0$, $a_z=0$ and $\phi=0$ on both sides of the simulation box] are always satisfied during the simulation. At $t=0.0T_0$ the laser pulse is apart from plasmas, and at later time, the laser irradiates on the plasmas. Our simulation is a global fluid-Maxwell simulation, which is quite different from previous studies that the soliton structure is put in as an initial condition \cite{Phys.Rev.E.65.016405,Phys.Rev.E.70.036403,PhyPla.13.074504,PhyPla.13.032309}.
The exact definition of a fluid implies the accumulation of a macroscopic number of particles within the interest region, and it seems that the fluid treatment of the region of zero plasma density may result in unphysical outcome. 
When numerically integrating the equation of motion Eqs.\ (5) and (6) within the vacuum-like region, 
the lack of the balance force from the electrostatic field may result in the unphysical $x$ direction speeding of the electrons and ions under the laser ponderomotive force. However, the plasma density is zero, and the total $x$ direction current is zero too, 
which does not contribute to the solving of the coupled Maxwell equations and 
makes the global fluid-Maxwell simulation applicable\cite{PhyPla.12.062308}.

To ensure the accuracy of the numerical scheme, we follow the energy of the system at each time step during our simulations. 
The total energy can be separated into four components: 
\begin{eqnarray}\label{10,11,12,13}
&& \epsilon_{l}=\frac{1}{2}\int dx[(\frac{\partial a_y}{\partial t})^2+(\frac{\partial a_z}{\partial t})^2+(\frac{\partial a_y}           {\partial x})^2+(\frac{\partial a_z}{\partial x})^2], \\
&& \epsilon_{p}=\frac{1}{2}\int dx[(\frac{\partial \phi}{\partial x})^2], \\
&& \epsilon_{e}=\int dx[(\gamma_e-1)n_e], \\
&& \epsilon_{i}=\int dx[m(\gamma_i-1)n_i],
\end{eqnarray}
where $\epsilon_{l}$ is the laser energy, $\epsilon_{p}$ is the energy of the plasma wave, $\epsilon_{e}$ is the electron kinetic energy and $\epsilon_{i}$ is the ion kinetic energy. In Eq.\ (10), we can see that the laser energy consists of four parts: $y$ component electric energy [$({\partial a_y}/{\partial t})^2$], $y$ component magnetic 
energy [$({\partial a_z}/{\partial x})^2$], $z$ component electric energy [$({\partial a_z}/{\partial t})^2$] and $z$ component magnetic energy [$({\partial a_y}/{\partial x})^2$]. It should be emphasized that this field is plagued by substandard simulation
methodology, which has led to bold conclusions being drawn from what amounts to numerical artifacts. The energy conserving of the 
simulations convince us that the new physics we report is not from numerical effect. 

We first study the problem of an ultra-short CP laser pulse penetration into over-dense plasmas. The parameters of the incident CP laser pulse are: normalized amplitude $a_y=0.71$, $a_z=0.71$, [$\sin(\pi t/2\tau)^2$] time profile with $\tau=5.0T_0$ and laser frequency $\omega_0=0.88$. The background plasma density is set to be $n_e=1.0$, corresponding to $n_e=1.3n_c$, in which $n_c=m_e\omega_0^2/4{\pi}e^2$. The vacuum-plasma is modeled with [$\sin(\pi x/2s)^2$] density profile, where $s$ is the shape factor
to describe the steepness of the vacuum-plasma interface. 

\begin{figure*}\label{1}
\includegraphics[width=16.0cm]{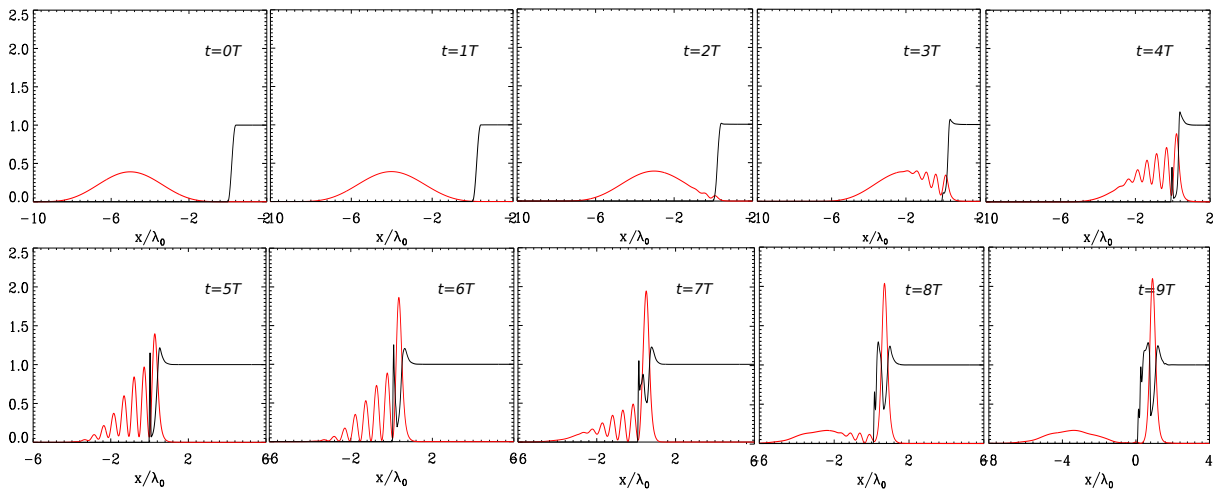}
\caption{\label{f1} (color online) The formation process of CP solitons: red line represents the normalized laser intensity and the black line is for the plasma density. Here, $a_y=0.71$, $a_z=0.71$, $\tau=5T_0$, $\omega_0=0.88$, $n_e=1.0$ and $s=0.4\lambda_0$.}
\end{figure*}

In Fig.\ 1, the space-time dynamics of the penetration process is demonstrated when there is a sharp density slope at the plasma boundary. A relativistic electromagnetic soliton is generated during the penetration process, which is in agreement with the theoretical prediction that the penetration occurs through soliton-like structure\cite{Phys.Rev.E.62.1234}. Compared with the incident pulse, the penetrated soliton owns a sharply narrow width which is less than one laser wavelength and the amplitude is greatly enhanced. After the formation, the soliton propagates into the uniform plasmas with a constant velocity $v=0.172$, which is shown in Fig.\ 2 (a). From Fig.\ 3, the energy history demonstrates that during the soliton formation stage,
within the early $7.0T_0$, there is an energy transfer from the incident laser to the background plasmas. It is easy to understand this energy transfer mechanism: the ponderomotive force drives the plasmas inward in the laser propagation direction, during this process work has been done by the laser pulse. However during the propagation process of the soliton, we see from Fig.\ 3 that the energy of the laser pulse and background plasmas do not change any more, which demonstrates that the CP soliton propagation is quasi-static. 

\begin{figure}\label{2}
\includegraphics[width=8.5cm]{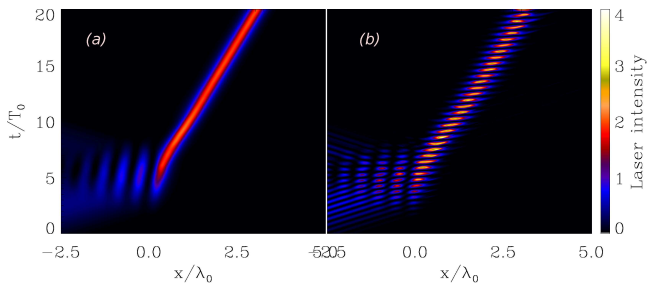}
\caption{\label{f2} (color online) The space-time distribution of the normalized laser intensity: (a) is for CP laser and (b) for LP laser. Here the simulation parameters for CP laser pulse are the same as that of Fig.\ 1, and the simulation parameters for LP laser pulse are $a_y=1.00$, $a_z=0.0$, $\tau=5T_0$, $\omega_0=0.88$, $n_e=1.0$ and $s=0.1\lambda_0$.}
\end{figure}

\begin{figure}\label{3}
\includegraphics[width=8.5cm]{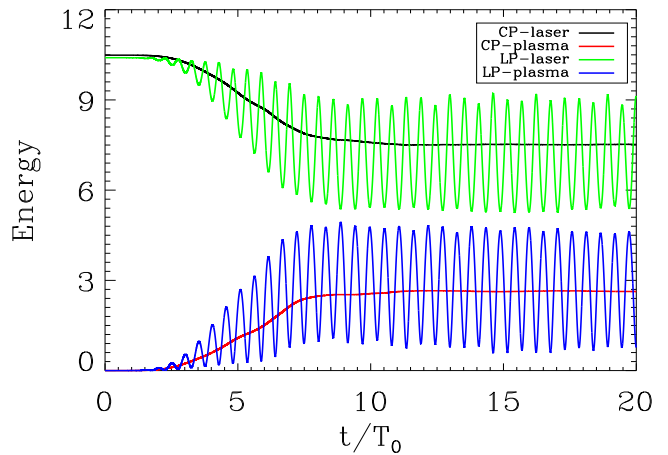}
\caption{\label{f3} (color online) The energy history of the fluid-Maxwell simulations: back line represents the normalized energy of the CP laser, the red line represents the plasma energy (mainly the electron kinetic energy) driven by CP laser, the green line is for energy of the LP laser and blue line is for the plasma energy by LP laser. Here the simulation parameters are the same as that of Fig.\ 2.}
\end{figure}

\begin{figure}\label{4}
\includegraphics[width=8.5cm]{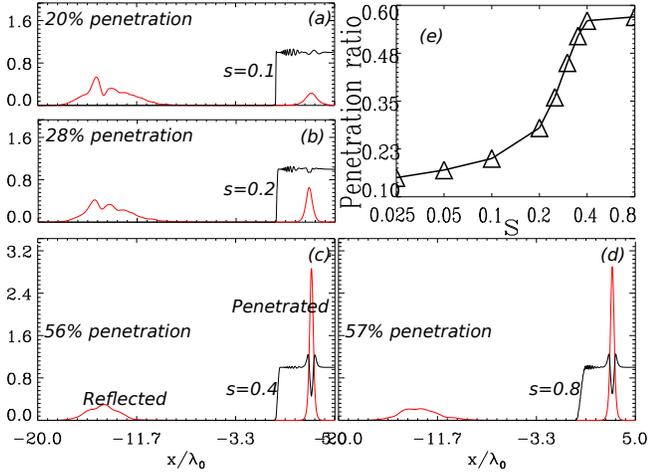}
\caption{\label{f4} (color online) (a)-(d) demonstrate the reflected and penetrated CP laser pulse structure under different vacuum-plasma steepness, where the red line is the laser intensity and black line is the plasma electron density. (e) The penetration ratio of the incident laser pulse vs. shape factor $s$, where the black triangle is from the fluid-Maxwell simulation data.}
\end{figure}

Considerable part of the incident CP laser pulse is transformed into the sharply narrow and relativistic intense electromagnetic soliton. Our simulations demonstrate that there is an obvious relation between the penetration ratio and the steepness of the vacuum-plasma interface. As is shown in Fig.\ 3 (a)-(d), for a relatively steep interface the penetrated CP soliton is quite weak, and the penetration ratio is rather low. In contrast, the rather smooth interface allows a strong penetration. Fig.\ 3 (e) shows the penetration ratio for different shape factor $s$, demonstrating the general penetration ratio range $10\% \sim 60\%$. 

Fig.\ 5 shows the spectral intensity of the initially incident CP irradiation, the reflected part and the penetrated soliton.
In the soliton formation stage, the ponderomotive force pushes the plasmas inward in the direction of the incident pulse propagation. As red line shows, the reflected radiation from the vacuum-plasma interface is mostly red-shifted. The frequency of the penetrated CP soliton is blue-shifted for $s=0.1$ [subplot (a) in Fig.\ 4] and red-shifted for $s=0.4$ [subplot (c) in Fig.\ 4]. The behavior of the weak penetrated soliton [$s=0.1$, subplot (a) in Fig.\ 4] can be well described by the known finite soliton velocity solutions, which are based on the weak density response\cite{PhysRevLett.68.3172}
\begin{eqnarray}
&& a_y=a_m \text{sech} [\frac{(1-v^2-\eta^2)^{1/2}}{1-v^2}(x-v t)]\times\cos[\frac{\eta v}{1-v^2}(x-\frac{t}{v})], \nonumber \\
&& \text{where} \nonumber \\
&& a_m=4[\frac{(1-v^2)(1-v^2-\eta^2)}{4\eta^2-(1-v^2)(3+v^2)}]^{1/2}\nonumber
\end{eqnarray}
is the amplitude of the CP soliton, and $\eta$ is the frequency parameter with $\eta=\omega(1-v^2)$. The carrier frequency of the weak penetrated soliton can be derived from the frequency-amplitude-velocity relation,
\begin{eqnarray}\label{14}
\omega=\frac{[16(1-v^2)^2+a_m^2(1-v^2)(3+v^2)]^{1/2}}{(1-v^2)[4 a_m^2+16(1-v^2)]^{1/2}}.
\end{eqnarray}
For the weak penetrated soliton [$s=0.1$, subplot (a) in Fig.\ 4] with $a_m=0.482$ and $v=0.172$, Eq.\ (14) gives $\omega=1.008=1.145\omega_0$, which is quite consistent with the one $1.142\omega_0$ obtained from the fluid-Maxwell simulation in Fig.\ 5 (a).  

To study the penetration process of the LP laser, we should be very careful. The density ripples caused by the second-harmonic oscillating ponderomotive force will break down the simulation process, while an fully implicit energy-conserving numerical scheme is applied to well deal with this problem. In addition, in order to avoid or neglect the vacuum-heating and the $\bm{J}\times\bm{B}$ effect on the vacuum-plasma interface, the incident laser pulse should be ultra-short to shorten the interaction time between the incident laser and the sharp vacuum-plasma boundary. We change the incident laser amplitude (the same intensity with that of CP) to $a_y=1.0$ and $a_z=0$, while keeping other parameters the same as that of CP laser. 
Similar to the CP case, during the early $7.0T_0$ interaction time, the energy is transfered from the incident LP laser pulse into background plasmas in one way although there is some small scale perturbations.
Fig.\ 2 (b) shows that the propagation velocity of the penetrated electromagnetic structure is almost the same as that of CP soliton which is $v=0.172$.  However, we find that unlike the CP soliton, during its propagation, there exist an energy transition mechanism between the LP penetrated structure and the background plasmas with a frequency about double the incident laser frequency. The energy exchange is even more clearly indicated from the energy history map shown in Fig.\ 3, where the laser 
energy and the background plasma energy coupled together and exchange with each other with a frequency $2\omega_0$. 

However, from Fig.\ 6 (c) and (d) for the LP penetration case, we can see that the penetrated electromagnetic structure disappears at $t=16.00T_0$ and suddenly reappears at $t=16.25T_0$, with a period $T_0/2$. 
At $t=16.00T_0$ almost all the penetrated electromagnetic energy is transformed into background plasmas. 
While at $t=16.25T_0$ the electromagnetic structure has returned back from the background plasmas. 
It behaves like a blinking breather, which is quite different from the quasi-stable soliton as shown in Figs.\ 6 (a) and (b).   

\begin{figure}\label{5}
\includegraphics[width=8.5cm]{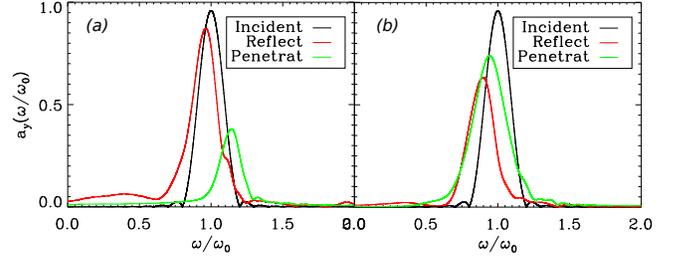}
\caption{\label{f5} (color online) The spectral intensity of the incident CP laser pulse, reflected part and penetrated part. The black line is for initially incident CP laser pulse, red line is for reflected and green line for penetrated. (a) is the case $s=0.1$ and (b) is the case $s=0.4$. Here the remaining simulation parameters are the same as that of Fig.\ 1.}
\end{figure}

Here, we provide a qualitative interpretation for the energy transition mechanism with $2\omega_0$ between the LP penetrated structure and the background plasmas. During our simulations, we follow the kinetic energy of the electrons, the kinetic energy of the ions, the laser energy and the plasma wave energy. As the electron kinetic energy [$\epsilon_{e}$] occupies the main part of the background plasma energy [$\epsilon_{i}+\epsilon_{e}+\epsilon_{p}$], the plasma wave energy does not dominate during these process, thus we can drop the Poisson Equation [Eq.\ (7)] during our analysis. We now focus on the Eq.\ (1). The refraction term from the background ion motions can be neglected due to the high ion to electron mass ratio and we set the electron momentum along the laser propagation direction to $p_e=0$, then Eq.\ (1) can be rewritten as
\begin{equation}\label{15} 
\frac{\partial^2 a_y}{\partial t^2}-\frac{\partial^2 a_y}{\partial x^2}+\frac{n_e}{(1+a_y^2)^{1/2}}a_y=0. \\ 
\end{equation}
As $a_y=a_m\sin(\omega_0 t)$, the refraction term $n_e/(1+a_y^2)^{1/2}$ which describes the effective plasma density under the LP laser irradiation becomes
$n_e/[1+a_m^2/2-a_m^2/2\cos(2\omega_0 t)]^{1/2}$. 
In our fluid-Maxwell simulations, the laser frequency is $\omega_0=0.88$ which corresponds to $n_c=0.76$ and the background plasma density is $n_e=1.0$ which corresponds to $n_e=1.3n_c$.
Unlike the soliton structure in the CP laser case where there is a density cavity for the background plasmas, in the LP laser case, the background density is almost unchanged, $n_e\sim1.0$, except for some small scale and fast oscillating density ripples. 
Taking into the simulation result, $a_m\sim2.0$, the effective plasma density 
$n_e/[1+a_m^2/2-a_m^2/2\cos(2\omega_0 t)]^{1/2}$
can range from $0.45$ to $1.00$ where the critical density $n_c$ located at $0.76$. Let us consider the two extreme conditions. 
If the effective plasma density reaches $0.45$, it means $\gamma_e$ reaches maximum or the background plasmas obtain its highest energy. However the low effective plasma density can not capture the relatively high frequency penetrated structure with $\omega_0=0.88$, and this penetrated electromagnetic structure just disperses away, as shown in Fig.\ 6 (c). 
If the effective plasma density arrives at $1.00$, the background plasmas reach its lowest energy. 
The relatively low frequency electromagnetic structure is well trapped by the high effective plasma density, 
which is shown in Fig.\ 6 (d). This process repeats at every half laser period as shown in Figs.\ 2 and 3. 

\begin{figure}\label{6}
\includegraphics[width=8.5cm]{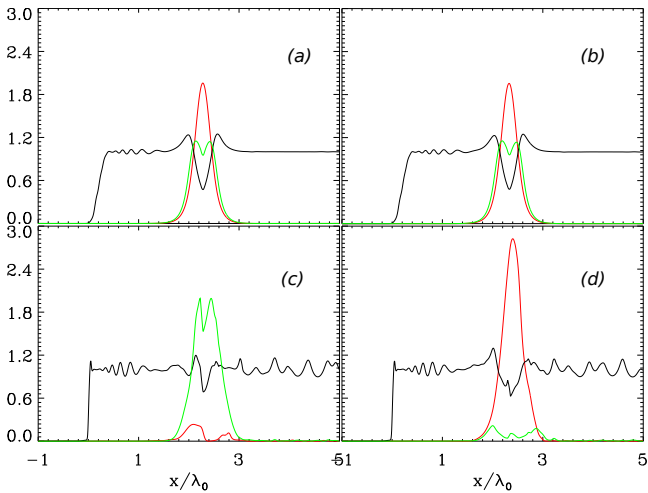}
\caption{\label{f6} (color online) The distribution of laser intensity and plasma density: (a) and (b) is for CP laser pulse at $t=16.00T_0$ and $t=16.25T_0$; (c) and (d) for LP laser pulse at $t=16.00T_0$ and $t=16.25T_0$. The black line represents the plasma electron density, red line is for laser intensity and green line for electron kinetic energy density. Here the simulation parameters are the same as that of Fig.\ 2.}
\end{figure}

When the background plasma density $n_e$ is greater than $1.5n_c$, 
the interaction results in the generation of plasma field structure consisting of alternating plasma and vacuum
regions with the CP laser penetrating into the over-dense plasmas over a finite length, which is consistent with the previous studies \cite{PhysRevLett.87.275002,PhyPla.12.062308}. Because the formation time of the alternating plasma field structure is quite long which is about hundreds of laser period, it call for quite long incident laser pulse. 
As analyzed, it beyonds the ability of fluid-Maxwell simulations in studying such long pulse LP laser interaction with over-dense plasmas.

In summary, the relativistic electromagnetic penetration behavior is reexamined in the relativistic transparency region,
which is modeled by the relativistic hydrodynamic equations coupled with the full system of Maxwell equations. 
We have studied the penetration behavior through ultra-short CP and LP laser pulses interaction with over-dense plasmas. 
It is shown that for the ultra-short CP laser penetration occurs through a soliton like behavior, which is quite consistent with the existing studies. The penetration ratio of the ultra-short CP irradiation is systematically studied with different steepness of the vacuum-plasma interface, which indicates that a rather smooth vacuum-plasma interface allows a strong penetration. The propagation of the CP soliton in homogeneous background plasmas is quasi-static. However, we have found that the ultra-short LP laser penetrates through a breather like behavior, and there is an energy transition mechanism with a frequency $2\omega_0$ between the penetrated breather like structure and the background plasmas. 
A qualitative interpretation has been given to describe this mechanism of energy exchange. 


\begin{acknowledgments}
This work was supported by the National Natural Science Foundation of
China (Grant Nos. 11075025, 10835003, 10905004, 11025523, and 10935002) and the Ministry of Science and Technology of China (Grant No. 2011GB105000).
\end{acknowledgments}


\begin{thebibliography}{99}
\bibitem{numerical} The numerical scheme is to be presented elsewhere.
\end{thebibliography}
{}

\end{document}